\documentclass[12pt]{article}
\usepackage{amsmath,amssymb}
\newtheorem{thm}{Theorem}[section]
\newtheorem{lemma}[thm]{Lemma}
\newtheorem{cor}[thm]{Corollary}
\newtheorem{prop}[thm]{Proposition}
\newtheorem{conj}{Conjecture}
\newcommand{\R}{{\mathbb R}}
\newcommand{\bp}{{\bar\partial}}
\newcommand{\bpt}{{\bar D}}
\newcommand{\G}{{\cal G}}

\newcommand{\C}{{\mathbb C}}
\newcommand{\Z}{{\mathbb Z}}
\newcommand{\geu}{{\gamma_{\text{Euler}}}}
\renewcommand{\H}{{\mathbb H}}
\renewcommand{\P}{{\mathbb P}}

\newcommand{\Res}{\text{Res}}
\newcommand{\eps}{\epsilon}

\renewcommand{\Im}{\mbox{Im}}
\renewcommand{\Re}{\mbox{Re}}
\newenvironment{proof}{
    \noindent{\bf Proof:} \hspace*{1em}}{
    \hspace*{\fill} $\square$\medskip }

\newcommand{\PSbox}[3]{\mbox{\rule{0in}{#3}\hspace{#2}\includegraphics{#1}}}

\begin{document}
\title{The Laplacian and $\bp$ operators on critical planar
graphs}
\author{Richard Kenyon
\thanks{Laboratoire de Math\'ematiques
UMR 8628 du CNRS, B\^at 425, Universit\'e Paris-Sud,
91405 Orsay, France.}}
\date{}
\maketitle
\abstract{
On a periodic planar graph whose edge weights satisfy
a certain simple geometric condition, the discrete Laplacian and
$\bp$ operators have the property that
their determinants and inverses only depend on
the local geometry of the graph.

We obtain explicit expressions for the logarithms of the
(normalized) determinants, as well as the inverses of these
operators. 

We relate the logarithm of the determinants to the
volume plus mean curvature of an associated hyperbolic ideal polyhedron.

In the associated dimer and spanning tree models, for which the
determinants of $\bp$ and the Laplacian respectively play the role of the
partition function, this allows us to compute the entropy and correlations 
in terms of the local geometry.

In addition, we define a continuous family of special discrete
analytic functions, which, via convolutions gives a
general process for constructing discrete analytic functions
and discrete harmonic functions on critical planar graphs.
}

\section{Introduction}
Let $\G$ be a graph and $c$ a nonnegative
weight function on the edges of $\G$ defining the edge conductances. 
We define 
the {\bf Laplacian operator} $\Delta\colon \C^{|\G|}\to \C^{|\G|}$ by
$$\Delta f(u) = \sum_{w\sim u} c(uw)(f(u)-f(w))$$
where the sum is over vertices $w$ adjacent to $u$.
This is one of the most basic and useful operators on $\G$.

On the superposition of $\G$
and its planar dual $\G^*$, we can define another operator,
the $\bp$ operator, which defines ``discrete
analytic functions'' (those which satisfy 
$\bp f=0$).
In fact we will define $\bp$ on more general bipartite planar
graphs, see section \ref{bpscn} below.
In the form we use here the definition
comes from a statistical mechanics model, the dimer model:
the determinant of the $\bp$
operator is the partition function of the dimer model 
(see section \ref{dimerscn}).

We show that under a certain geometric condition, called isoradiality,
on the embedding of a planar graph, the local structure of
these two operators $\Delta$ and $\bp$ takes on a surprising simplicity.
In particular one can write down explicitly the determinants
and inverses of these operators, 
which only depend on local quantities:
the determinants only depend on the edge weights (not the combinatorics)
and the value of the inverse on two vertices $v_1,v_2$ only depends
on a path between the vertices.

In particular for the determinants we have the following results.
For the definitions see below.
\begin{thm}\label{Ztrees}
Let $\G_T$ be an isoradial embedding of a periodic planar graph and $c$ 
the associated weight function (giving the conductances on the edges).
Then the logarithm of the normalized determinant
of the Laplacian is
$$\log \rm{det}_1\Delta = \frac1{N}\sum_{\text{edges } e} 
\frac{2}\pi\left(L(\theta)+L(\frac{\pi}2-\theta)\right)+
\frac{2\theta}{\pi}\log\tan\theta,$$
where $N$ is the number of vertices in a fundamental domain, the 
sum is over the set of edges in a fundamental domain,
$c(e)=\tan(\theta(e))$ and
$L$ is the Lobachevsky function,
\begin{equation}\label{Lob}L(x)=-\int_0^x\log2\sin tdt.
\end{equation}
\end{thm}

The determinant of the Laplacian is useful since it computes
the partition function for several other statistical mechancial models,
among them the spanning tree model (see e.g. \cite{Biggs})
and the discrete Gaussian free field.
For the $\bp$ operator we have
\begin{thm}\label{Zdimers}
Let $\G_D$ be an isoradial embedding of a periodic bipartite
planar graph and $\nu$ the associated
weight function on edges. Then the normalized determinant of the 
discrete $\bp$ operator satisfies
$$\log\rm{det}_1\bp=
\frac1N\sum_{\text{edges } e}\frac1{\pi}L(\theta) + \frac{\theta}{\pi}\log 
2\sin\theta,$$
where $\nu(e)=2\sin\theta(e)$,
$N$ is the number of vertices in a fundamental domain, and the sum is over the 
edges in a fundamental domain.
\end{thm}

For results on and explicit expressions for the inverses of these operators, 
see Theorems \ref{CF} and \ref{green} below.

Similar statements can be made for nonperiodic graphs, 
on condition that the right-hand sides makes sense; for this
one needs a unique ergodicity assumption on the translation
action on the graph. We will restrict ourselves to periodic cases here.

To a graph $\G_D$ with critical weight function $\nu$
as in Theorem \ref{Zdimers} is naturally associated
a $3$-dimensional ideal hyperbolic polyhedron (with infinitely
many vertices): it is the polyhedron $P$ 
in the upper-half space model of
$\H^3$ whose vertices are those of $\G_D^*$, the dual graph of $\G_D$.
We give a geometric realization
of a discrete analytic function as a deformation
of the associated embedding of the graph, alternatively as a deformation
of $P$.  Furthermore there is a relation between the
above determinants and the volume of the polyhedra, which identifies
the entropy of the corresponding dimer model to the volume,
and the mean energy to the mean curvature.

These results were inspired by the work of and discussions with
Christian Mercat \cite{Mercat}
who used isoradial embeddings (``critical'' embeddings in his terminology)
in the study of discrete analytic functions and the Ising model. 
Isoradial embeddings apparently first appeared in the work of Duffin
\cite{Duffin}. Our definition of ``elementary'' harmonic functions, see 
section \ref{discretscn}, also appeared independently in
\cite{merc2}, where a more detailed study of these functions is undertaken.

The outline of the paper is as follows.
\begin{itemize}\item
Section \ref{critscn}. We define ``isoradial'' embeddings
and critical weight functions for planar graphs. 
\item
Section \ref{bpscn}. We define the $\bp$ operator,
discrete analytic functions, and show in what sense they give 
perturbations of the embedding. We also prove the existence,
unicity and asymptotics of the inverse $\bp^{-1}$,
as well as giving an explicit local formula. 
\item Section \ref{detscn}. We define the normalized determinant $\det_1\bp$ 
and prove Theorem \ref{Zdimers}. 
\item Section \ref{Lapscn}.
We point out the relation between the Laplacian and our $\bp$ operator
in the case of graphs $\G_T$.
\item Section \ref{greenscn}.
We compute the Green's function on $\G_T$ and its asymptotics. 
\item Section \ref{cvxscn}. We prove that the
set of isoradial embeddings of a graph is convex (and usually
infinite dimensional), and $\log\det_1\Delta$ and
$\log\det_1\bp$ are concave functions on it. 
\item Section \ref{discretscn}. We discuss a general construction
of discrete analytic and discrete harmonic functions on critical
graphs. 
\item Section \ref{dimerscn}. We discuss connections
between the $\bp$ operator and the dimer model. Here we relate
the volume and mean curvature
of the polyhedron $P$ with the entropy and mean energy of the dimer model.
\end{itemize}

\section{Critical weight functions}\label{critscn}
\subsection{Polyhedral embeddings}
Let $\G$ be a planar graph, and $\G^*$ its planar dual.
For a vertex $v\in \G$ we denote $v^*$ its dual face of $\G^*$. 
Similarly for an edge $e$ we denote by $e^*$ its dual edge.

A {\bf polyhedral embedding} of $\G$
is an embedding of $\G$ in the plane
such that every face is a cyclic polygon, that is,
is inscribable in a circle. A {\bf regular polyhedral embedding}
is a polyhedral embedding in which the circumcenters of the faces
are contained in the closures of the faces. 

A polyhedral embedding of a graph $\G$ defines a
map from $\G^*$, the dual of $\G$, to the plane, 
by sending a dual vertex to the circumcenter of the corresponding face
of $\G$, and a dual edge to
the segment joining its endpoints. 
If the embedding of $\G$ is regular, this map will be
an embedding of $\G^*$ unless 
some circumcenters are on the boundary of their faces, in which
case some dual vertices may have the same image in the plane.
In other words some edges of $\G^*$ may have length $0$.

The term polyhedral refers to the fact that a polyhedral
embedding gives rise to a three-dimensional ideal hyperbolic
polyhedron, in the following way. The upper half-space model of hyperbolic
space $\H^3$ is identified with $\{(x,y,z)\in\R^3|z>0\}$.
A polyhedral embedding of $\G$ on the $xy$-plane is the vertical projection
of the edges of an (infinite) ideal polyhedron $P(\G)$ whose vertices
are the vertices of $\G$. 
If the polyhedral embedding is regular then $P(\G)$ is convex;
the dihedral angle at an edge is the angle of 
intersection of the circumcircles
of the two adjacent faces.
This polyhedron will play a role in section \ref{dimerscn}.

\subsection{Isoradial embeddings}
An {\bf isoradial} embedding is a regular polyhedral embedding
in which all circumcircles have the same radius.

\begin{figure}[htbp]
\vskip3in
\caption{\label{GG}Isoradial embedding of $\G$ (thick lines) 
and associated rhombi (thin lines).}
\PSbox{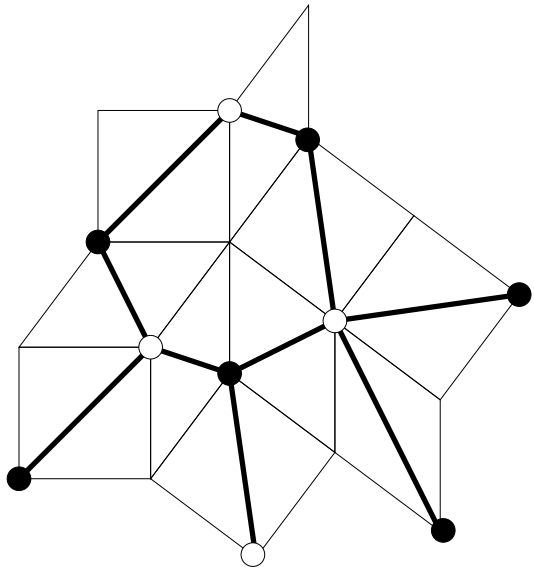}{0in}{0in}
\end{figure}

If the embedding of $\G$ is isoradial then the embedding
of $\G^*$ will also be isoradial, with the same radius.
We will always take the common radius to be $1$.
To each edge $e$ of $\G$
we associate a rhombus $R(e)$
whose vertices are the vertices of $e$ and the vertices of its dual edge.
The rhombus is therefore of edge length $1$ and
the angle at a vertex of $\G$ is in $[0,\pi]$.
See Figure \ref{GG}. 

A {\bf chain} of rhombi is an bi-infinite sequence of rhombi 
$\{\dots,x_i,x_{i+1},\dots\}$
with $x_{i+1}$ adjacent to $x_i$ along the edge opposite
the edge along which $x_{i-1}$ and $x_i$ are adjacent.
The rhombi in a chain all contain an edge parallel to a given direction.
When $\G$ is bipartite, the vertices of $\G$ in a rhombus chain
are alternately black and white, so that a chain has a ``black'' side and
a ``white'' side.

Note that a rhombus chain cannot cross itself (since it is monotone)
and two rhombus chains cannot cross each other more than once,
since at the crossing point the rhombus
which both chains have in common has orientation determined by
the relative orientations of the two chains.

Which planar graphs have isoradial embeddings? 
Here is an explicit conjecture. A {\bf zig-zag} path in a planar
graph is an edge path which alternately turns right and left, 
that is, upon arriving at a vertex it takes the first edge to the left,
then upon arriving at the next vertex takes the first edge to the right,
and so on. If the graph has an isoradial embedding a zig-zag
path corresponds to a chain of rhombi. Consequently two necessary
conditions for the existence of an isoradial embedding are that a zig-zag
path cannot cross itself, and two zig-zag paths cannot cross each other
more than once. We conjecture that these two conditions are 
sufficient as well. 

\begin{conj} 
A planar graph has an isoradial embedding if and only if
no zig-zag path crosses itself, and no two zig-zag paths cross
each other more than once.
\end{conj}

\subsection{Critical weights}
Suppose we have a isoradial embedding of $\G$.
For each edge $e$ of $\G$ define $\nu(e)=2\sin\theta$
and $c(e)=\tan\theta$,
where $2\theta$ is the angle of the rhombus $R(e)$ at the vertices
it has in common with $e$. We refer to $\theta$ as the
{\bf half-angle} of $R$.
The function $\nu$ is the weight
function for the $\bp$ matrix and $c$
is the weight function (conductance) for the Laplacian matrix.
A weight function $\nu$ or $c$
which arises in this way is called {\bf critical}.
Note that $\nu(e)$ is the length of $e^*$, the dual edge of $e$.

Note that a critical weight function determines
the embedding of $\G$ and $\G^*$.

It is useful to allow the rhombus half-angles to be in $[0,\pi/2]$,
that is, to allow degenerate rhombi with half-angles $0$ or $\pi/2$. 
In this case the
embeddings of $\G$ or $\G^*$ can be degenerate in the sense that
two or more vertices may map to the same point. 

If the half-angles 
of the rhombi in a chain are in $[0,\pi/2]$ but bounded away from $0$
and $\pi/2$  we can deform the angles of all the rhombi in the chain,
simply by changing slightly the direction of the common parallel edge.
The new rhombus angles 
define a new critical weight function. Moreover one
can show that every critical weight function can be obtained
from any other by deforming in this way along chains: see section
\ref{cvxscn}.

\subsection{Periodic graphs}

For the rest of the paper we suppose that $\G$ is an infinite {\bf periodic}
graph, that is, the embedding (and therefore the weight function)
is invariant by translates by a two-dimensional lattice $\Lambda$ in $\R^2$,
and the quotient $\G/\Lambda$ is finite. 
We denote this quotient $\G_1$.

We use periodicity in a fundamental way in defining 
and proving the formulas for the determinants, Theorems \ref{Ztrees}
and \ref{Zdimers}.
For the inverses of the operators, however, periodicity is not necessary.
As long as the graph has a reasonable regularity, for example if
there are only a finite number of rhombus angles, then 
the results on the inverses extend (with essentially no modifications
to the proofs). 

\section{The $\bp$ operator}\label{bpscn}
\subsection{Definition}
Let $\G_D$ be a bipartite planar graph, that is, the vertices of $\G_D$ can be
divided into two subsets $B\cup W$ and vertices in $B$ (the black vertices)
are only adjacent to vertices in $W$ (the white vertices) and vice versa.
Suppose that $\G_D$ has an isoradial embedding.
Let $\nu$ be a critical weight function on $\G_D$.

We define a symmetric matrix $\bp$ indexed by the vertices of $\G_D$ as follows.
If $v_1$ and $v_2$ are not adjacent $\bp(v_1,v_2)=0$.
If $w$ and $b$ are adjacent vertices, $w$ being white and $b$ black,
then $\bp(w,b)=\bp(b,w)$ is the complex number of
length $\nu(wb)$ and direction pointing from $w$ to $b$.
If $w$ and $b$ have the same image in the plane then
the $|\bp(w,b)|=2$, and the 
direction of $\bp(w,b)$ is that which is perpendicular to 
the corresponding dual edge (which has nonzero length by definition),
and has sign determined by the local orientation. Another useful way to say
this is as follows. If rhombus $R(wb)$
has vertices $w,x,b,y$ in counterclockwise (cclw) order then 
$\bp(w,b)$ is $i$ times the complex vector $x-y$.

The matrix $\bp$ defines the {\bf $\bp$ operator}.
The $\bp$ matrix is also
called a {\bf Kasteleyn matrix} for the underlying dimer model, 
see section \ref{dimerscn}.
Since $\bp$ maps $\C^B$ to $\C^W$ and $\C^W$ to $\C^B$,
it really consists of two operators $\bp_{BW}\colon\C^B\to\C^W$
and its transpose $\bp_{WB}\colon\C^W\to\C^B.$

As an example, note that when $\G_D=\Z^2$ with edge lengths $r=\sqrt{2}$
the $\bp$ operator has a more recognizable form,
$$\bp f(v) := \sqrt{2}\left(f(v+r)-f(v-r)+if(v+ri)-if(v-ri)\right).$$

\subsection{Discrete analytic functions and perturbations}
\label{pertsection}
A function $f\colon B\to\C$ satisfying $\bp f\equiv 0$, that is
$\sum_{b\in B} \bp(w,b)f(b)=0$ for all $w$,
is called a {\bf discrete analytic function}. This generalizes the
definition given in \cite{Mercat}. Similarly a function $f\colon W\to\C$ 
satisfying $\bp f\equiv 0$ is also called a discrete analytic function.
So a critical bipartite planar graph supports two independent families
of discrete analytic functions.

The simplest example of a discrete analytic function is a constant
function (as the reader may check).

\begin{figure}[htbp]
\vskip4in
\PSbox{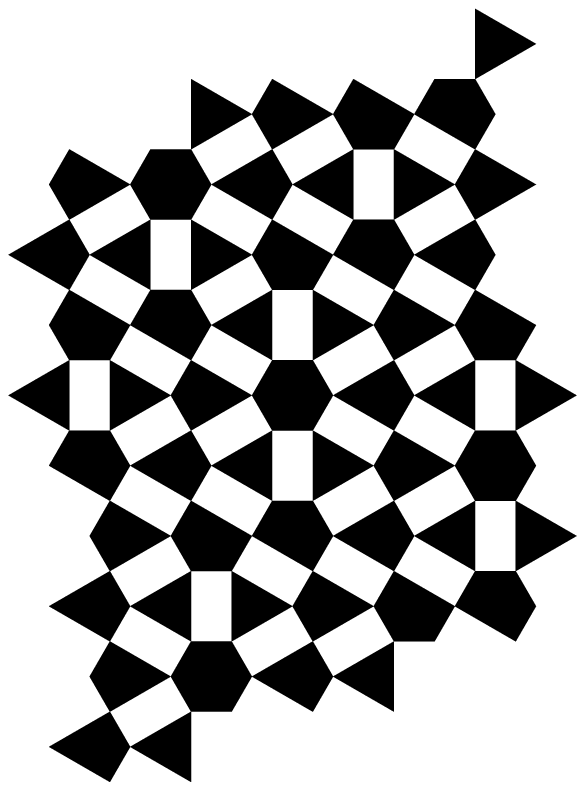}{0in}{0in}
\PSbox{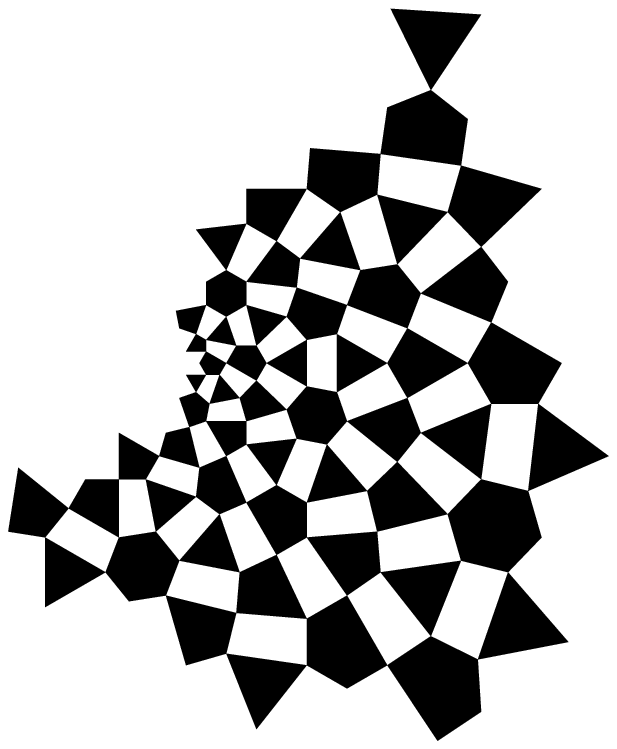}{3in}{0in}
\caption{\label{pertn}A discrete analytic function is a perturbation
of the embedding of $\G_D$. This example is a special case where $\G_D$ is
a superposition of an isoradial graph and its dual, see
section \ref{Lapscn}.}
\end{figure}

A discrete analytic function $F\colon B\to\C$
defines a perturbation of the embedding of $\G_D^*$,
as follows (this idea generalizes an example in \cite{CdV}). 
The perturbation is a complex homothety 
(rotation, scaling and translation) on each black
face but deforms the white faces in a more general way.
See Figure \ref{pertn}.

The perturbation is defined as follows. Fix a small $\eps>0$.
To each face $b^*$ of $\G_D^*$
we apply the map $z\mapsto z(1+\eps F(b))+ t_b$,
where $t_b$ is a certain translation. The $t_b$ are chosen so that 
if two black faces meet at a vertex, their images still meet at the 
(image of the) same vertex.
This is of course only possible if for each white face
the polygonal path formed by the translates of the images of its edges 
is still a closed polygon: 
let $w$ be a white
vertex of $\G_D$, and $b_1,\dots,b_m$ its neighboring black vertices. 
Let $e^*_1,\dots,e^*_m$ be the edges running cclw
around the face $w^*$ of $\G_D^*$,
so that $e^*_j$ is the dual edge to $wb_j$. Because the edges and dual edges are
perpendicular, as a complex vector we have $e_j^*=i\bp(w,b_j)$.
Then $\bp F=0$ implies that
$$\bp(w,b_1)F(b_1)+\dots+\bp(w,b_m)F(b_m)=0,$$ 
or
\begin{equation} \label{ee}
e^*_1 F(b_1)+\dots+e^*_m F(b_m)=0
\end{equation}
and since $e_1^*+\dots+e_m^*=0$, we have
\begin{equation} \label{ee2}
e^*_1(1+\eps F(b_1))+\dots+e^*_m(1+\eps F(b_m))=0.
\end{equation}
In conclusion we can find $t_b$ so that 
each white face is again a closed polygon.

\begin{figure}[htbp]
\vskip5in
\PSbox{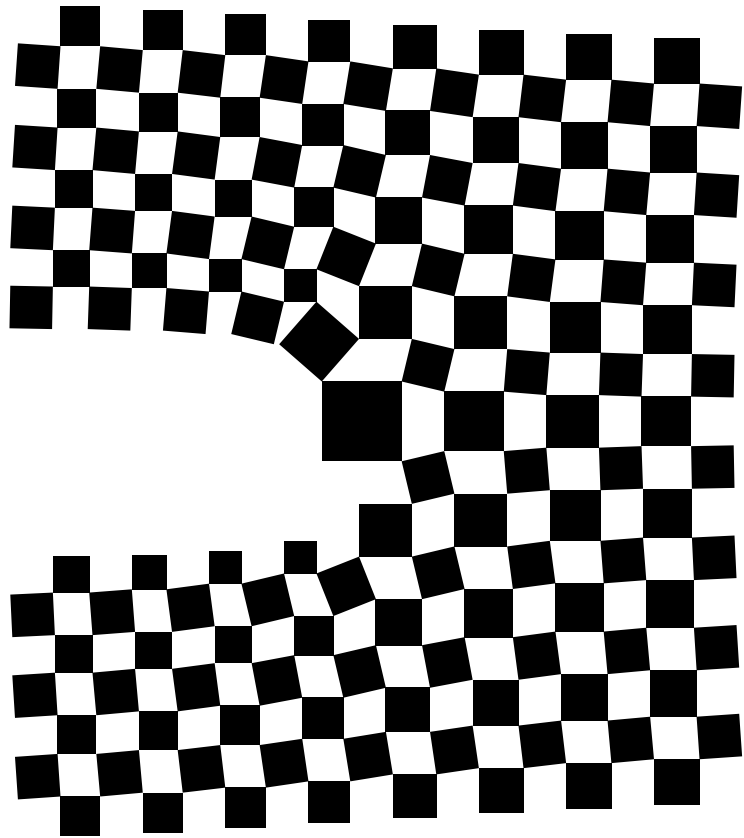}{0in}{0in}
\caption{\label{sqlog}
The function $\bp^{-1}(w,\cdot)$ on the graph $\Z^2$ defines
a perturbation of the standard embedding (here we took the 
$\eps$ of (\ref{ee2}) to be $3.5$).}
\end{figure}

The inverse $\bp^{-1}(w,b)$ which we define in the next section 
is a discrete analytic function of $b$ for each fixed $w$,
except for a single singularity:
it satisfies 
$$\sum_{b\in B}\bp(w',b)\bp^{-1}(w,b)=\delta_{w'}(w)=\left\{
\begin{array}{ll}0&\mbox{if }w\neq w'\\1&\mbox{if }w=w'\end{array}\right..$$
Therefore for a fixed $w$ it defines a perturbation of the embedding
except that the polygon $w^*$ is not closed:
at the face $w^*$ the right-hand side of (\ref{ee}) is $1$ not $0$. 
So we can therefore define only a map from the 
branched cover of $\G_D^*$, branched around $w$,
to $\C$; it has a non-trivial holonomy around $w$. 
See Figure \ref{sqlog} for $\bp^{-1}$ in the case $\G_D=\Z^2$.

\section{The $\bp^{-1}$ operator}
For critical bipartite graphs $\G_D$ we define $\bp^{-1}$ 
to be the unique function satisfying
\begin{enumerate}
\item $\bp\bp^{-1}=Id$
\item $\bp^{-1}(w,b)\to 0$ when $|b-w|\to\infty$.
\end{enumerate}

We first show unicity, then existence in the following section.

\subsection{Unicity}\label{infperiodic}
\begin{thm} There is at most one $\bp^{-1}$ satisfying the above 
two properties.
\end{thm}
\begin{proof}
We first show that for all $w,b$ the argument of $\bp^{-1}(w,b)$ is 
well defined up to additive multiples of $\pi$. 
To do this we conjugate $\bp$ by a diagonal unitary matrix to make it real.
This conjugation simply involves multiplying the 
edge weights on edges incident to each vertex 
by a constant (a constant depending on the vertex).

Fix a vertex $v_0$ of $\G_D$.
Let $v$ be any other vertex of $\G_D$ and
$v_0,v_1,\dots,v_{2k}=v$ be a path of rhombus edges from $v_0$ to $v$.
Each edge $v_jv_{j+1}$ has exactly one vertex of $\G_D$
(the other is a vertex of $\G_D^*$). Direct the edge
away from this vertex if it is white, and towards this vertex if it is black.
Let $e^{i\alpha_j}$ be the corresponding vector (which may point contrary
to the direction of the path from $v_0$ to $v$).
We multiply all the edges incident to $v$ with weight
$s_v=\pm e^{-\frac{i}2\sum\alpha_j}$
(either choice of sign will do). Note that these multipliers
are well-defined up to $\pm1$: 
the product of the multipliers on a path surrounding
a rhombus is $\pm1$. Let $S$ be the diagonal matrix of
these weights (with an arbitrary choice of signs).
Then we claim that
each entry of $S^*\bp S$ is real: if $w,b$ are adjacent in $\G_D$,
and $e^{i\theta},e^{i\phi}$ are the edge vectors of the rhombus edge $R(wb)$, 
so that $\bp(w,b)=i(-e^{i\theta}+e^{i\phi}),$ then
$$S^*\bp S(w,b)=s_w^*s_b \bp (w,b)= 
\pm s_w^*(s_w e^{-i(\theta+\phi)/2})(i(-e^{i\theta}+e^{i\phi}))\in\R.$$

We now claim that $S^*\bp S$ has at most one inverse tending to zero
at infinity. If $F_1,F_2$ are two such inverses,
then $F_1-F_2$ satisfies $S^*\bp S(F_1-F_2)\equiv 0$.
Moreover by reality of $S^*\bp S$ we can suppose $F_1-F_2$ is real
(otherwise take the real or imaginary parts separately).

Let $w_0$ be a fixed white vertex of $\G_D$.
As in section \ref{pertsection} we associate to each black 
vertex $b$ a polygon similar to the face $b^*$ of $\G_D^*$. That is we
associate a translate of $[S(F_1-F_2)S^*(w_0,b)]\cdot b^*$.
These translates are chosen so that if polygons $b_1^*,b_2^*$
meet at a vertex then their images meet at the the image of the same vertex.
For each white face $w^*$ of $\G_D^*$,
the images of its edges all have the same direction: 
if $b$ is a neighbor of $w$ then the dual edge $(wb)^*=e^{i\theta}-e^{i\phi}$ 
is mapped to the vector
$$(e^{i\theta}-e^{i\phi})S(F_1-F_2)S^*(b,w_0)=
(e^{i\theta}-e^{i\phi})(s_w e^{-i(\theta+\phi)/2})s_{w_0}^*,$$
which is $s_{w_0}^*s_w$ times a real number. Hence up to $\pm 1$
the direction is independent of the choice of neighbor $b$ of $w$.
The image of these edges is therefore a segment. 
Thus we have a map from $\G_D^*$ to the plane, such that
each black face is mapped to a similar copy of itself (positively oriented)
and each white face is mapped to a segment. Each such segment
has a neighborhood in $\R^2$ of its interior covered by the images of its
neighboring black faces.
If $(F_1-F_2)(b,w_0)$ is non-zero at some $b$, 
the image of this map is open, since the image extends
across the boundary of each black polygon.
But this is impossible if $F_1-F_2$ tends to zero at $\infty$.
\end{proof}

\subsection{Existence}
For a critical weight function $\nu$ on a bipartite graph $\G_D$
we have the following explicit description of $\bp^{-1}$.

Let $w_0$ be a white vertex.
Define for every vertex $v$ a rational function $f_v(z)$ as follows.
Let $w_0=v_0,v_1,v_2,\dots,v_k=v$ be a path in the rhombus tiling
from $w_0$ to $v$.
Each edge $v_jv_{j+1}$ has exactly one vertex of $\G_D$
(the other is a vertex of $\G_D^*$). Direct the edge
away from this vertex if it is white, and towards this vertex if it is black.
Let $e^{i\alpha_j}$ be the corresponding vector (which may point contrary
to the direction of the path).
We define $f_v$ inductively along the path, starting from 
\begin{equation}\label{cond1}
f_{v_0}=1.
\end{equation}
If the edge $v_jv_{j+1}$ leads away from a white vertex or towards
a black vertex, then 
\begin{equation}\label{propagate1}
f_{v_{j+1}}=\frac{f_{v_j}}{(z-e^{i\alpha_j})},\end{equation}
else if it leads towards a white or away from a black, then
\begin{equation}\label{propagate2}
f_{v_{j+1}}=f_{v_j}\cdot(z-e^{i\alpha_j}).\end{equation}
To see that $f_v(z)$ is well-defined, it suffices to show that
the multipliers for a path around a rhombus come out to $1$.
For a rhombus $R(wb)$ with vertices $w,x,b,y$ such that the vector
$wx$ is $e^{i\theta}$ and the vector $wy$ is $e^{i\phi}$,
we have 
$$f_w(z)=f_y(z)(z-e^{i\phi})=f_b(z)(z-e^{i\theta})(z-e^{i\phi})=
f_x(z)(z-e^{i\theta}).$$
This shows that $f$ is well-defined.

Any function $f_v(z)$ satisfying (\ref{propagate1}),(\ref{propagate2})
(but not necessarily (\ref{cond1})) is a discrete analytic function for $\G_D$,
see Theorem \ref{specialdafs} below.
We call such functions {\bf special} discrete analytic functions.

For a black vertex $b$ the coupling function 
$\bp^{-1}(w_0,b)$ is defined roughly
as the sum over the poles of $f_b(z)$
of the residue of $f_b$ times the angle of $z$ at the pole. 
However there is an ambiguity in the choice
of angle, which is only defined up to a multiple of $2\pi$.

To make this definition precise, we have to assign angles (in $\R$,
not in $[0,2\pi)$) to the poles of $f_b(z)$. 
We work on the branched cover of the plane, branched over $w_0$,
so that for each black vertex 
$b$ in this cover we can assign a real angle $\theta_0$
to the complex vector $b-w_0$, 
which increases by $2\pi$ when $b$ winds once around $w_0$.
Each pole of $f_b$ corresponds to a rhombus chain separating $b$ from
$w_0$, with $b$ being on the black side of the chain. 
Because the chain is monotone we can assign an angle to the common
parallel (recall that this points from the white side to the
black side of the chain) which is in $(\theta_0-\pi,\theta_0+\pi)$.
This assigns a real angle to each pole of $f_b$.

\begin{thm}\label{CF}
We have 
\begin{equation}
\label{Kinverse}
\bp^{-1}(w_0,b)= 
-\frac{1}{2\pi i}\sum_{\text{poles } e^{i\theta}} \theta\cdot
\Res_{z=e^{i\theta}}(f_b),
\end{equation}
where the angles $\theta\in\R$ are chosen as above for some lift of $b$.
This can be written
$$\frac1{4\pi^2i}\int_C f_b(z)\log z dz,$$
where $C$ is a closed contour surrounding cclw the part of the circle
$\{e^{i\theta}~|~\theta\in[\theta_0-\pi+\eps,\theta_0+\pi-\eps]\}$
which contains all the poles of $f_b$,
and with the origin in its exterior.
\end{thm}

\begin{proof} 
Let $F(b)$ denote the right-hand side of (\ref{Kinverse}). 
Note that if we choose a different
lift of $b$, the angles all change by a constant multiple of $2\pi$;
however the sum of all residues of $f$ is zero since $f_b(z)$ has a double
zero at $\infty$. Therefore the right-hand side is independent of the
lift of $b$.

We need to show that
$\sum_{b\in B}\bp(w,b)F(b) = \delta_{w_0}(w)$, and
$F(b)$ tends to zero when $b\to\infty$ (see section \ref{infperiodic}).

Take a white vertex $w$, $w\neq w_0$, and 
let $b_1,\dots,b_k$ be the neighbors of $w$.
Let $e^{i\theta_j},e^{i\phi_j}$ be the edges of $R(w,b_j)$, so that
$\bp(w,b_j)=i(-e^{i\theta_j}+e^{i\phi_j})$.
For each pole $e^{i\theta}$ of some $f_{b_j}$, the value of the angle $\theta$
is the same for each of the $b_j$ (they are part of the same chain
for which $e^{i\theta}$ is the common parallel). 
Let $C_{\theta}$ be a small loop around $e^{i\theta}$ in $\C$.
We have
\begin{eqnarray*}
\sum_{b\in B}\bp(w,b)F(b)
&=& \sum_{j=1}^ki(-e^{i\theta_j}+e^{i\phi_j})F(b_j),\\
&=&-\frac{1}{2\pi i}\sum_{j=1}^ki(-e^{i\theta_j}+e^{i\phi_j})
\sum_{\text{poles }e^{i\theta}} 
\theta\cdot \Res_{z=e^{i\theta}}(f_{b_j})\\
&=&-\frac1{2\pi}\sum_{\text{poles }e^{i\theta}}\theta\cdot\sum_{j=1}^k
(-e^{i\theta_j}+e^{i\phi_j})\frac1{2\pi i}\int_{C_\theta} f_{b_j}(z)dz\\
&=&-\frac1{2\pi}\sum_{\text{poles }e^{i\theta}}\frac{\theta}{2\pi i}
\int_{C_\theta} f_{w}(z)
\sum_{j=1}^k
\frac{(-e^{i\theta_j}+e^{i\phi_j})}{(z-e^{i\theta_j})(z-e^{i\phi_j})}dz\\
&=&-\frac1{2\pi}\sum_{\text{poles }e^{i\theta}}\frac{\theta}{2\pi i}
\int_{C_\theta} f_{w}(z)\left(\sum_{j=1}^k
\frac1{(z-e^{i\phi_j})}-\frac1{(z-e^{i\theta_j})}\right)dz\\
&=&-
\frac1{2\pi}\sum_{\text{poles }e^{i\theta}}\frac{\theta}{2\pi i}
\int_{C_\theta} 0 dz = 0.
\end{eqnarray*}

However when $w=w_0$ we have 
$$F(b_j)=-\frac1{2\pi i}\left(\frac{\theta_j-\phi_j}{e^{i\theta_j}-e^{i\phi_j}}
\right),$$
so that 
$$\sum_{j=1}^ki(-e^{i\theta_j}+e^{i\phi_j})F(b_j)=\frac1{2\pi}\sum_{j=1}^k
\theta_j-\phi_j = 1,$$
since the angles increase by $2\pi$ around $w_0$.

Finally to show that $F(b)$ tends to zero see Theorem \ref{1/z} below.
\end{proof}

\subsection{Asymptotics of $\bp^{-1}$}
Let $w=v_0,v_1,\dots,v_{2k}=b$ be a path of rhombus edges
from $w$ to $b$ which crosses each rhombus chain at most once.
Such a path exists since two chains cross each other at most once.
Every vertex $v_{2j}$ is a vertex of $\G_D$, and 
the edges coming to and leaving $v_{2j}$ contribute
a pole and a zero to $f_b$ (we are assuming $f_w(z)=1$). 
So $f_b(z)$ has the form 
$$f_b(z)=\frac1{(z-e^{i\theta_1})(z-e^{i\theta_2})} 
\prod_{j=1}^{k-1} \frac{(z-e^{i\alpha_j})}{(z-e^{i\beta_j})}.$$
By periodicity of $\G_D$, the angles $\beta_j$ are all in 
$[\theta_0-\pi+\eps,\theta_0+\pi-\eps]$ for some fixed $\eps>0$
independent of $w,b$. 

\begin{thm}\label{1/z}
We have 
$$\bp^{-1}(w,b) = \frac{1}{2\pi}\left(\frac{1}{b-w}+
\frac{\gamma}{(\bar b-\bar w)} \right)+
\frac1{2\pi}\left(\frac{\xi_2}{(b-w)^3}+\frac{\bar\xi_2}{\gamma
(\bar b-\bar w)^3}\right)+
O\left(\frac1{|b-w|^{3}}\right),$$
where $\gamma=f_b(0)=e^{-i(\theta_1+\theta_2)}\prod e^{i(-\beta_j+\alpha_j)}$ 
and $\xi_2=e^{2i\theta_1}+e^{2i\theta_2}+\sum e^{2i\beta_j}-e^{2i\alpha_j}$
in the above notation.
\end{thm}

\begin{proof}
By Theorem \ref{CF}, $\bp^{-1}(w,b)$ is given by an integral 
$$\frac1{(2\pi)^2i}\int_C f_b(z)\log z dz$$ 
on a simple closed curve $C$ which surrounds in a cclw sense the set
$\{e^{i\theta}~:~\theta\in[\theta_0-\pi+\eps,\theta_0+\pi-\eps]\}$
and has the origin 
in its exterior. In fact since $f$ tends to zero at $\infty$
we can homotope $C$ to the curve from $\infty$ to the origin
and back to $\infty$ along the two sides of the ray $R$ from the origin
in direction $\theta_0+\pi$. 
On the two sides of
this ray, $\log z$ differs by $2\pi i$: it is $2\pi i$ less on the cclw
side, and the integral therefore
becomes 
$$\frac1{2\pi}\int_R f_b(z) dz,$$ 
the integral being from $\infty$ to $0$ along $R$.

We show that when $|b-w|$ is large the only contributions 
to this integral come from a
neighborhood of the origin and from a neighborhood of $\infty$.
Suppose without loss of generality that $\theta_0=0$, so that
$R$ is the negative real axis. 
Define
$$N=|b-w|=
|e^{i\theta_1}+e^{i\theta_2}+\sum_{j=1}^{k-1} e^{i\beta_j}-e^{i\alpha_j}|.$$

For small $t$ we have 
$$\frac{t-e^{i\alpha}}{t-e^{i\beta}}=\frac{e^{i\alpha}}{e^{i\beta}}\cdot
\exp\left((e^{-i\beta}-e^{-i\alpha})t+(e^{-2i\beta}-e^{-2i\alpha})t^2/2+O(t^3)
\right),$$
giving
\begin{equation}\label{ap1}
f_b(t) = \gamma
\exp\left((\bar b-\bar w)t+ \overline{\xi_2}\frac{t^2}2+O(t^3)\right).
\end{equation}

Since $\beta\in[-\pi+\eps,\pi-\eps]$, we have $\cos\beta>1-\delta$
for some fixed $\delta=\delta(\eps)$.
For all $t<0$ we can use the bound
\begin{eqnarray*}
\left|\frac{t-e^{i\alpha}}{t-e^{i\beta}}\right|&=&
\sqrt{\frac{t^2+1-2t\cos\alpha}{t^2+1-2t\cos\beta}}\\
&\leq&
\sqrt{1+\frac{2t(\cos\beta-\cos\alpha)}{t^2+1-2t(1-\delta)}}\\
&\leq&
\exp\left(\frac{t(\cos\beta-\cos\alpha)}{t^2+1-2t(1-\delta)}\right),
\end{eqnarray*}
which gives 
\begin{equation}\label{ap2}
|f_b(t)|\leq \frac{e^{\frac{t}{t^2+1-2t(1-\delta)}
\sum \cos\beta_j-\cos\alpha_j}}{
|(t-e^{i\theta_1})(t-e^{i\theta_2})|}\leq
e^{t(N-2)/(1+t^2-2t(1-\delta))}.
\end{equation}

For large $t$ we have 
$$\frac{t-e^{i\alpha}}{t-e^{i\beta}}=\frac{1-e^{i\alpha}/t}{1-e^{i\beta}/t}=
\exp\left((e^{i\beta}-e^{i\alpha})\frac1t+(e^{2i\beta}-e^{2i\alpha})
\frac{1}{2t^2}
+O(\frac1{t^3})\right),$$
which gives
\begin{equation}\label{ap3}
f_b(t)= e^{(b-w)/t+\xi_2/(2t^2)+O(t^{-3})}.
\end{equation}

When $-1/\sqrt{N}\leq t\leq 0$, we can use the approximation (\ref{ap1}),
when $- N^{1/2}\leq t\leq 1/\sqrt{N}$ we can use (\ref{ap2})
and when $t\leq -N^{1/2}$ we can use (\ref{ap3}).

The small-$|t|$ integral gives
$$\frac{\gamma}{2\pi} \int_{-1/\sqrt{N}}^0 e^{t(\bar b-\bar w)+
\bar\xi_2t^2/2+O(Nt^3)}dt=
\frac{\gamma}{2\pi}\left(\frac1{\bar b-\bar w}+\frac{\bar\xi_2}{
(\bar b-\bar w)^3}\right)+O(\frac1{N^3}).$$
The intermediate integral is negligible, and the large-$t$ integral
gives
$$\frac1{2\pi}\int_{-\infty}^{-N^{1/2}}\frac1{t^2}
e^{(b-w)/t+\xi_2/2t^2+O(N/t^3)}dt=
\frac1{2\pi}\left(\frac1{b-w}+\frac{\xi_2}{(b-w)^3}\right)+
O(\frac{1}{N^3}).$$
\end{proof}

\section{Determinants}\label{detscn}
For a finite graph $\G$ we define the {\bf normalized determinant}
$\det_1 M$ 
of an operator $M\colon\C^{\G}\to\C^{\G}$ to be 
$$\mbox{det}_1 M\stackrel{\text{def}}{=}|\det M|^{1/|\G|},$$ where
$|\G|$ is the number of vertices of $\G$.

For an operator $M$ on a finite graph,
$\det M$ is a function of the matrix entries $M(i,j)$
which is linear in each entry separately.
In particular for an edge $e=ij$, 
as a function of the matrix entry 
$M(i,j)$ we have $\det M=\alpha+\beta M(i,j)$, 
where $\beta$ is $(-1)^{i+j}$ times
the determinant of the minor obtained by removing
row $i$ and column $j$. 
That is, 
$$\frac{\partial(\det M)}{\partial(M(i,j))}= M^{-1}(j,i)\cdot\det M,$$
or 
\begin{equation}\label{detdef}
\frac{\partial(\log\det M)}{\partial M(i,j)}= M^{-1}(j,i). 
\end{equation}

Suppose that $\G$ is periodic
under translates by a lattice $\Lambda$.
Let $\G_n = \G/n\Lambda$, the finite graph which is the
quotient of $\G$ by $n\Lambda$. Now if we sum (\ref{detdef}) for all
$\Lambda$-translates of edge $ij$ in $\G_n$, and then divide
both sides by $|\G_n|$, it yields
\begin{equation}\label{det1def}
\frac{\partial(\log\det_1 M)}{\partial M(i,j)}= \frac1{|\G_1|}M^{-1}(j,i),
\end{equation}
where $M(i,j)$ is now the common weight of all translates of edge $ij$,
that is, the left-hand side is the change in $\log\det_1 M$ when
the weight of all translates of $ij$ changes.
Note that this equation is independent of $n$.

We can now use (\ref{det1def})  
to {\it define} the normalized determinant of $\bp$ on an infinite 
periodic graph $\G_D$ in terms of $\bp^{-1}$.
Rather, it defines the change in the normalized determinant.
To complete the definition we need to define the normalized determinant of the 
$\bp$ operator at some initial condition.
If the half-angles of a graph degenerate by tending to $0$ or $\pi/2$,
the edge weights tend to either $0$ or $2$.
The edges with weight $0$ contribute nothing,
the edge of weight $2$ each are defined to contribute $\sqrt{2}$ to the 
normalized determinant.
Thus we define the limit of $\log\det_1\bp$ to be $\frac{n}{2|\G_1|}\log 2$,
where $n$ is the number of edges per fundamental domain 
with weight $2$.
This makes sense since under this degeneration, removing the zero
weight edges, the graph consists of
copies of $\Z$, each consisting of weight $2$ edges.

\medskip

\noindent{\bf Proof of Theorem \ref{Zdimers}:\hskip.5cm }
Suppose $\G_D$ is periodic with lattice $\Lambda$.
Take a rhombus chain and all its $\Lambda$-translates
in $\G_D$. Let $\phi_0$ be the smallest
half-angle of a rhombus on these chains, and $\phi_1$ the largest 
half-angle: they exist by periodicity. 
If $\phi_0\leq\pi/2-\phi_1$, deform
these chains by changing the common parallels 
to decrease $\phi_0$ to zero: at that point the 
corresponding edge weight is zero and
we can remove the edge completely from $\G_D$, simplifying
the graph in the process.
If on the other hand $\pi/2-\phi_1<\phi_0$, deform the chain by increasing
$\phi_1$ to $\pi/2$. 
At this point the rhombus $R_1$
of angle $\phi_1$ has flattened out to become a segment.
We now repeat with another chain, but in future deformations
we always keep $R_1$ as a segment: the angle $\phi_1$ remains $\pi/2$.
We need to show that this is always possible.
If we deform along a chain or union of chains
whose common parallels do not contain
an edge of $R_1$, the deformation will not affect $R_1$.
If an edge of $R_1$ is in the common parallel of a chain being
deformed, we deform simultaneously and in the same direction 
{\it both} chains which pass through $R_1$. So a deformation
involves changing the common angle of parallels of chains, all of
which have a parallel in the same direction.
We can deform in this way because for any set of
chains having parallels in the same
direction, there is a deformation which tilts all the common parallels
and is a translation on each complementary component of these chains.
We continue deforming until all angles are zero or $\pi/2$. 
We have proved that any rhombus tiling of the plane can be deformed
through rhombus tilings to flatten out all the rhombi: the rhombus vertices
all map to $\Z$, the rhombus edges map to edges of $\Z$, and the
edges of $\G_D$ either have length $2$ (and weight zero) or length $0$
(and weight $2$). 

We can compute the change
in $\log\det_1\bp$ as in (\ref{det1def})
under such a deformation since it only depends on the
angles of the rhombi in the chains.

To complete the calculation it suffices to compute the contribution 
for a deformation along a single chain. 
In fact we can treat each rhombus of the chain separately.
Suppose we deform a chain whose common parallel is $e^{2i\theta}$.
Suppose the rhombus $R(wb)$ has edges $e^{i\alpha}$ and $e^{i(\alpha+2\theta)}$
leading away from $w$.   Then the term in 
$$|\G_1|\frac{d\log\det_1\bp}{d\theta}$$ which involves
this rhombus is
\begin{eqnarray*}
|\G_1|\frac{d(\log\det_1\bp)}{d(\bp(w,b))}\cdot\frac{d(\bp(w,b))}{d\theta}&=&
\bp^{-1}(w,b)\frac{d(\bp(w,b))}{d\theta}\\
&=&
-\frac1{2\pi i}\left(\frac{2\theta}{e^{i(\alpha+2\theta)}-e^{i\alpha}}\right)
\frac{d}{d\theta}(i(-e^{i(\alpha+2\theta)}+e^{i\alpha}))\\
&=&\frac{\theta e^{i\theta}}{\pi\sin\theta}.
\end{eqnarray*}
Suppose in the above deformations this rhombus has half-angle going to zero. 
Integrating the real part of this for $\theta$ from $0$ to $\phi$ 
yields 
$$\frac1{\pi}\int_0^{\phi}\theta\cot\theta d\theta=
\frac{\phi}{\pi}\log\sin\phi-
\frac1{\pi}\int_0^{\phi}\log\sin\theta d\theta$$
(integrating by parts), which is
$$=\frac{\phi}{\pi}\log2\sin\phi+\frac1{\pi}L(\phi).$$

On the other hand if the rhombus half-angle goes to $\pi/2$, 
we should integrate from $\pi$ down to $\phi$. 
This yields 
$$-\frac12\log2+\frac{\phi}{\pi}\log2\sin\phi+\frac1{\pi}L(\phi),$$ 
since when $\phi=\pi/2$ we have
$$\frac{\phi}{\pi}\log2\sin\phi+\frac1{\pi}L(\phi)=\frac12\log 2.$$
This completes the proof.
\hspace*{\fill} $\square$\medskip 

Note that for $\Z^2$ in its standard embedding this gives
the classical result of Kasteleyn \cite{Kast1,CKP}, 
except for an extra factor
$\frac12\log 2$ per edge since our edge weights are $\nu(e)=\sqrt{2}$,
not $1$.

A more classical way of defining the normalized determinant
is via an exhaustion of $\G_D$ by finite graphs.
On an infinite graph $\G_\infty$ which is a limit of a growing sequence
of finite graphs $\G_\infty=\lim_{n\to\infty}\G_n$, and supposing
that an operator $M_\infty$ on $\G_\infty$ is a limit
in an appropriate sense of $M_n$ on $G_n$, one would wish to define
$\det_1 M_\infty \stackrel{def}{=} \lim_{n\to\infty} \det_1 M_n,$
assuming this limit exists.
Unfortunately it is not so easy to prove that the limit
exists in the case of the operator $\bp$, even when the graph
is $\Z^2$ with periodic weights: see \cite{CKP}. That is why we chose the
previous definition instead.

\section{Relating the Laplacian to $\bp$}\label{Lapscn}
Let $\G_T$ be a planar graph (not necessarily bipartite)
with an isoradial embedding in the plane.
To an edge $e$ with rhombus angle $2\theta\in [0,\pi]$ we associate
a weight $c(e)=\tan\theta$. This weight $c(e)$ is the conductance
in the underlying electrical network.
This is the situation considered in \cite{Mercat}.

\begin{figure}[htbp]
\vskip3in
\PSbox{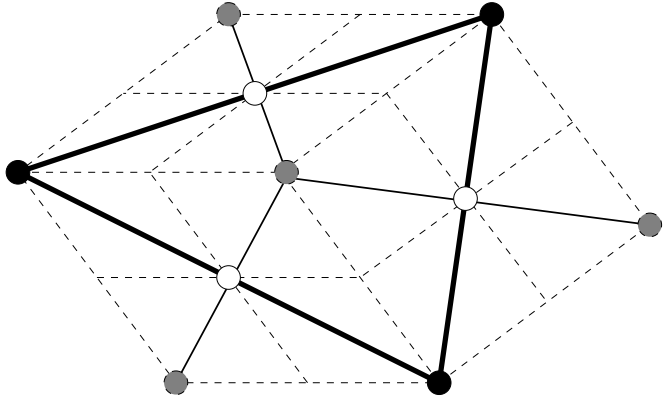}{0in}{0in}
\caption{\label{superposition}
Isoradial embedding of $\G_T$ (black vertices, thick lines) and its
dual $\G_T^*$ (grey vertices, thin lines), and associated rhombi 
of $\G_D$ (dotted lines).}
\end{figure}

Given $\G_T$ we define another graph $\G_D$ to be 
the superposition of $\G_T$ and its dual, that is, $\G_D$ has a vertex
for each vertex, edge and face of $\G_T$, and an edge for each half-edge
of $\G_T$ and each half dual-edge as well. See Figure \ref{superposition}.
The graph $\G_D$ is a bipartite planar graph, with black
vertices of two types: vertices of $\G_T$ and vertices of $\G_T^*$,
and white vertices of $\G_D$ correspond to the edges of $\G_T$. 

An isoradial embedding of $\G_T$ gives rise to an isoradial embedding of
$\G_D$: each rhombus of $\G_T$ with half-angle $\theta$
is divided into four congruent rhombi
in $\G_D$, two of half-angle $\theta$ and two of half-angle 
$\frac{\pi}2-\theta$:
see Figure \ref{superposition}. 

An edge of $\G_T$ of rhombus half-angle $\theta$ has weight $c(e)=\tan\theta$.
This  defines weight $\tan\theta$ on the 
two ``halves'' of each edge in $\G_T\cup \G_T^*$ (which are edges of 
$G_D$), and the dual edges are naturally weighted $1$ (as in \cite{KPW}). 

The $\bp$ operator on $\G_D$ and the Laplacian on $\G_T$ 
are related as follows.  Let $\bpt$ be obtained from $\bp$
by multiplying edge weights of $\G_D$
around each white vertex (coming from an edge of $\G_T$ of half-angle $\theta$)
by $1/(2\sqrt{\sin\theta\cos\theta})$, that is,
$\bpt(w,b)=S\bp(w,b)S^*$ where $S$ is the diagonal matrix
defined by $S(w,w)=1/(2\sqrt{\sin\theta(w)\cos\theta(w)})$ and $S(b,b)=1$.

Then an edge of $\G_D$ coming
from a ``primal'' edge of $\G_T$ has
weight $\sqrt{\tan\theta}$ for $\bpt$, 
and an edge coming from a dual edge
has weight $1/\sqrt{\tan\theta}$ for $\bpt$. (These edges have
weight $2\sin\theta,2\cos\theta$ respectively for $\bp$.)

\begin{lemma} \label{KK=Del}
Restricted to vertices of $\G_T$
we have $\bpt^*\bpt=\Delta_{\G_T}.$
Restricted to faces of $\G_T$, we have
$\bpt^*\bpt=\Delta_{\G_T^*},$ the Laplacian on $\G_T^*$.
\end{lemma}

\begin{proof}
Let $b\in \G_T$ have neighbors $b_1,\dots,b_k$ and let $\theta_i$ be 
the half-angle of edge $bb_i$. Then in $\G_D$, $b$ has neighbors $w_1,\dots
w_k$ with these same half-angles, and $\bpt(b,w_j)=\alpha_j\sqrt{\tan\theta_j}$
where $|\alpha_j|=1$.

We have
$$\bpt^*\bpt(b,b) = \sum_{j=1}^k |\alpha_j\sqrt{\tan\theta_j}|^2
= \sum_{j=1}^k \tan\theta_j=\Delta(b,b).$$
If $b,b'$ are adjacent in $\G_T$ we have
$$\bpt^*\bpt(b,b')=
(\alpha\sqrt{\tan\theta})(-\bar\alpha\sqrt{\tan\theta})=-\tan\theta
=\Delta(b,b').$$
If $b,b'$ are at distance $2$ in $\G_D$ but correspond to a vertex and
a face of $\G_T$, we have
two contributions to $\bpt^*\bpt(b,b')$ 
which cancel: in $\G_T$
let $e_1,e_2$ be the two edges from $b$ bounding the face $b'$, with 
$e_1$ on the right.
Then 
\begin{eqnarray*}
\bpt^*\bpt(b,b')&=&\bpt(b,w_1)\bpt^*(w_1,b')+\bpt(b,w_2)\bpt^*(w_2,b')\\
&=&(\alpha_1\sqrt{\tan\theta_1})(\frac{-i\overline{\alpha_1}}{
\sqrt{\tan\theta_1}})
+ (\alpha_2\sqrt{\tan\theta_2})(\frac{i\overline{\alpha_2}}{
\sqrt{\tan\theta_2}})\\
&=&0.
\end{eqnarray*}
Finally, if $b'$ is at distance at least $2$ in $\G_T$, 
$\bpt^*\bpt(b,b')=0$.

The proof for $\G_T^*$ is identical.
\end{proof}

\section{Green's function}\label{greenscn}
As we did for $\bp^{-1}$ we can give an explicit expression for the 
Green's function on a critical graph.

The Green's function $G(v,w)$ is defined to be the function satisfying 
\begin{enumerate}
\item $\Delta G(v,w)=\delta_v(w)$ (the Laplacian is taken with respect
to the second variable), 
\item $G(v,w)=O(\log(|v-w|)),$ and
\item $G(v,v)=0$.
\end{enumerate}

We define for each vertex $v$ of $\G_T$ and $\G_T^*$
a rational function $g_v(z)$ of $z$ as follows.
We define $g_v$ inductively starting from $g_{v_0}=\frac1{z}$
and if $v,v'$ are adjacent vertices of a rhombus then
\begin{equation}\label{harmprop}
g_{v'}=g_v\cdot\frac{z+e^{i\theta}}{z-e^{i\theta}},
\end{equation}
where $e^{i\theta}$ is the complex vector from $v$ to $v'$.
Now $g_v$ is well-defined since the product of the multipliers
around a rhombus is $1$.

\begin{thm}\label{green}
$$G(v_0,v_1)=
\frac{1}{4\pi i}\sum_{\text{poles } e^{i\theta}\neq 0} \theta\cdot
\Res_{z=e^{i\theta}}(g_{v_1}(z))=-\frac{1}{8\pi^2i}\int_C g_{v_1}(z)\log zdz,$$
where the angles $\theta$ and curve $C$ are chosen as in Theorem \ref{CF}.
For $v_0\in\G_T$ and $v_1\in\G_T^*$ this formula defines $i$ times the
harmonic conjugate of $G$.
\end{thm}

\begin{proof}
The proof is similar to the proof of Theorem \ref{CF}.
Clearly $G(v,v)=0$ by definition.
To show that $\Delta G(v,w)=\delta_v(w)$,
we show that for each edge 
the discrete Cauchy-Riemann equations are satisfied (\cite{Mercat}),
that is, we show that
for an edge $vv'$ with half-angle $\theta$ and with dual edge $ww'$, we have
$$(G(v_0,v')-G(v_0,v))i\tan\theta=G(v_0,w')-G(v_0,w),$$
assuming $ww'$ is obtained by turning cclw from $vv'$.
It suffices that for each pole $e^{i\phi}$ we have
\begin{equation}\label{gtan}
(g_{v'}(z)-g_v(z))i\tan\theta=g_{w'}(z)-g_{w}(z).
\end{equation}
By (\ref{harmprop}), we have 
$$g_{w'}(z)=g_v(z)\frac{z+e^{i\alpha}}{z-e^{i\alpha}}, \hskip1cm g_{w}(z)=
g_v(z)\frac{z+e^{i\beta}}{z-e^{i\beta}},\hskip1cm
g_{v'}(z)=g_v(z)
\frac{(z+e^{i\alpha})(z+e^{i\beta})}{(z-e^{i\alpha})(z-e^{i\beta})},$$
so that (\ref{gtan}) is 
$$g_v(z) \left(\frac{z+e^{i\alpha}}{z-e^{i\alpha}}
\frac{z+e^{i\beta}}{z-e^{i\beta}}-1\right)i\tan\frac{\alpha-\beta}2=
g_v(z)\left(
\frac{z+e^{i\alpha}}{z-e^{i\alpha}}-\frac{z+e^{i\beta}}{z-e^{i\beta}}\right).$$
This identity holds, therefore $G(v_0,w)$ is harmonic for $w\neq v_0$. 
Moreover for a dual vertex $v^*$ adjacent to $v_0$ we
have 
$$g_{v^*}(z)=\frac{z+e^{i\alpha}}{z(z-e^{i\alpha})},$$
whose residue at $e^{i\alpha}$ is $2$. Thus the value of $G(v_0,v)$
is $-\alpha i/2\pi$.
In particular for one cclw turn around the vertex $v_0$, the dual
function increases by $-i$, which implies
that the Laplacian at $v_0$ is $1$.
Finally to show that $G(v,w)=O(\log(|v-w|)),$ see Theorem \ref{Ggrowth} below.
\end{proof}

From Lemma \ref{KK=Del} there is a relation between $G$ and $\bpt^{-1}$:
\begin{cor} When edge $b_1b_1'$ of $\G_T$ corresponds to vertex $w_1$
of $\G_D$, we have
\begin{equation}\label{K=G-G}
\frac1{\bpt^*(w_1,b_1)}\bpt^{-1}(w_1,b_2)=
G(b_1,b_2)-G(b_1',b_2),\end{equation}
\end{cor}

\begin{proof}
By Lemma \ref{KK=Del},  taking the Laplacian of
$\bpt^{-1}(w_1,v)$ with respect to $v$ gives
\begin{eqnarray*}
\Delta_{\G_T}\bpt^{-1}(w_1,v)&=&\bpt^*\bpt\bpt^{-1}(w_1,v)\\
&=&\bpt^*\delta_{w_1}(v)=\bpt^*(w_1,b_1)
\left(\delta_{b_1}(v)-\delta_{b_1'}(v)+\frac{i}{\tan\theta}(
\delta_{f_1}(v)-i\delta_{f_2}(v))\right),
\end{eqnarray*}
where $f_1,f_2$ are the two faces adjacent to $e_1$.
In particular $\bpt^{-1}(w_1,b)$ is harmonic as a function
of $b\in \G_T$ except at $b_1$ and $b_1'$, where its Laplacian is
$\pm \bpt^*(w_1,b_1)$ respectively.

Since $\bpt^{-1}(w_1,b_2)\to0$ as $b_2\to\infty$
this proves (\ref{K=G-G}).
\end{proof}

\subsection{Asymptotics of the Green's function}
\begin{thm}\label{Ggrowth}
$$G(v_1,v_2)=-\frac1{2\pi}\log|v_2-v_1|-\frac{\geu}{2\pi}+
O(\frac1{|v_2-v_1|}).$$
\end{thm}

Note that our Laplacian is positive semidefinite, which results in
the minus sign in this formula.

\begin{proof}
Let $v_0=w_0,w_1,\dots,w_k=v_1$ be a path of rhombus edges from $v_0$ to
$v_1$. Let $e^{i\theta_j}$ be the complex vector from $w_j$ to $w_{j+1}$.
As in the proof of Theorem \ref{1/z} we 
assume without loss of generality that
$\theta_j\in[-\pi+\eps,\pi-\eps]$ for some fixed $\eps$ independent of 
$v_1,v_2$.

Let $N=|v_1-v_0|$.
We take $r\ll 1/N$ and $R\gg N$. 
We integrate
$$-\frac1{8\pi^2 i}\int_C \frac1z\prod_{j=0}^{k-1}
\frac{z+e^{i\theta_j}}{z-e^{i\theta_j}}\log zdz,$$ 
where $C$ is a curve
which runs cclw around the ball of radius $R$ around the origin,
from angle $-\pi$ to $\pi$, then along the negative $x$-axis
from $-R$ to $-r$, then clockwise around the ball of radius $r$ 
from angle $\pi$ to $-\pi$, and then back along the $x$-axis from
$-r$ to $-R$.

The integral around the ball of radius $r$ is 
$$-\frac1{8\pi^2 i}\int_{\pi}^{-\pi}(-1)^k(1+O(Nr))(\log r+i\theta)id\theta
=\frac{(-1)^k\log r}{4\pi}(1+O(Nr)).$$
The difference between the value of $\log z$ above and below the $x$-axis
is $2\pi i$, so that we can combine the two parts of the integral
along the $x$-axis into the integral of 
$$-\frac1{4\pi}\int_{-R}^{-r}\frac1z\prod_{j=0}^{k-1}
\frac{z+e^{i\theta_j}}{z-e^{i\theta_j}}dz.$$
This integral we split into a part from $-R$ to $-\sqrt{N}$,
from $-\sqrt{N}$ to $-1/\sqrt{N}$ and from $-1/\sqrt{N}$ to $-r$.

For small $t<0$ we have
$$\prod_{j=0}^{k-1}\frac{t+e^{i\theta_j}}{t-e^{i\theta_j}}=
(-1)^ke^{2\sum e^{-i\theta_j}t+O(t^2)}=(-1)^ke^{2(\bar v_1-\bar v_0)t+O(t^2)}$$
so that the integral near the origin is
$$-\frac{(-1)^k}{4\pi}
\int_{-1/\sqrt{N}}^{-r}\frac{e^{\bar\alpha t+O(t^2)}}t dt $$
where $\alpha=2(v_1-v_0),$
\begin{eqnarray*}&=&-\frac{(-1)^k}{4\pi}(1+O(\frac1N))
\int_{-\bar\alpha/\sqrt{N}}^{-\bar\alpha r}\frac{e^s}{s}ds\\
&=&-\frac{(-1)^k}{4\pi}(1+O(\frac1N))\left(
\int_{-\bar\alpha/\sqrt{N}}^{-1}\frac{e^s}{s}ds+
\int_{-1}^{-\bar\alpha r}\frac{e^s-1}{s}ds+\int_{-1}^{-\bar\alpha r}\frac{ds}s
\right)\\
&=&-\frac{(-1)^k}{4\pi}\left(\log(r\bar\alpha)+\geu\right)+O(\frac1N).
\end{eqnarray*}
The Euler $\gamma$ constant $\geu$ 
comes from evaluating the first two of these integrals
in the limit $\bar\alpha r\to0$ and $\bar\alpha/\sqrt{N}\to\infty$.

In the intermediate range the integral is negligible (see the
proof of Theorem \ref{1/z}).
Near $t=-R$ we have
$$\prod\frac{t+e^{i\theta_j}}{t-e^{i\theta_j}}=
e^{2\sum e^{i\theta_j}/t+O(t^{-2})}=e^{2(v_1-v_0)/t+O(t^{-2})}$$
so that the integral near $-R$ is
\begin{eqnarray*}
-\frac{1}{4\pi}\int_{-R}^{-\sqrt{N}}\frac{e^{\alpha/t+O(t^{-2})}}{t}dt
&=&-\frac{1}{4\pi}\int_{-R/\alpha}^{-\sqrt{N}/\alpha}\frac{e^{1/s}}{s}ds,\\
&=&-\frac{1}{4\pi}\left(
\int_{-R/\alpha}^{-1}\frac{e^{1/s}-1}{s}ds+\int_{-R/\alpha}^{-1}\frac{ds}s
+ \int_{-1}^{-\sqrt{N}/\alpha}\frac{e^{1/s}}{s}\right)\\
&=&-\frac1{4\pi}\left(-\log(R/\alpha)+\geu\right)+O(\frac1N),
\end{eqnarray*}
These are precisely the same integrals as in the ``near $r$'' case,
under the change of variable $s\to 1/s$.

Finally the integral around the ball of radius $R$ gives $-\frac1{4\pi}\log R$.
The sum is therefore
$$-\frac1{4\pi}(\log\alpha+(-1)^k\log\bar\alpha)-(1+(-1)^k)\frac{\geu}{4\pi}+
O(\frac1N)$$
which when $k$ is even is 
$$-\frac1{2\pi}\log|v_1-v_0|-\frac{\geu}{2\pi}+O(\frac1N)=
-\frac1{2\pi}\Re(v_1-v_0)- \frac{\geu}{2\pi}+O(\frac1N)$$
and when $k$ is odd it is 
$$-\frac1{4\pi}\log\frac{v_1-v_0}{\bar v_1-\bar v_0}+O(\frac1N)=
-\frac{i}{2\pi}\Im\log(v_1-v_0)+O(\frac1N).$$
\end{proof}

\subsection{Determinant of Laplacian}
Using Lemma \ref{KK=Del} we can compute the 
determinant of $\Delta$ in terms of the determinant of $\bp$.
However it is as simple to compute it directly.

We define the normalized determinant $\det_1\Delta$
to be the function of the conductances satisfying, for $w$ adjacent to $v$,
$$\frac{\partial(\log\det_1\Delta)}{\partial(\Delta(v,w))}=\frac1{|\G_1|}
\left(G(v,v)-G(v,w)-G(w,v)+G(w,w)\right)=-\frac{2G(v,w)}{|\G_1|},$$ 
and for a graph with zero conductances, $\det_1\Delta:=1$.

Let $\G_T$ be a periodic planar graph, periodic under
$\Lambda$ with fundamental domain $\G_T/\Lambda=\G_1$. 

\noindent{\bf Proof of Theorem \ref{Ztrees}}
The method of proof is the same as that of Theorem \ref{Zdimers}.
Take a chain of rhombi and all its translates under 
$\Lambda$, and change their common parallel,
decreasing the smallest angle to zero.
If an edge has angle which goes to zero, its conductance
goes to $0$ as well and so it can be removed from $\G_T$.
Similarly if any rhombus angle increases to $\pi/2$,
the corresponding conductance becomes infinite 
and one can remove the corresponding
edge, gluing the two vertices together.
In this way we can simplify the graph until no edges remain.

However there is a problem which is that the determinant blows up when
a conductance goes to $\infty$. So instead we compute 
the change in the difference
$${\cal E}:=
\log\text{det}_{1}\Delta - 
\frac1{|\G_1|}\sum_{e\in\G_1}\frac{2\theta}{\pi}\log\tan\theta$$
as we reduce the graph in this manner. (The quantity 
$\cal E$ is the entropy of the underlying spanning tree model.)

To compute the change under the perturbations we can
treat each rhombus independently.
For a rhombus $R(vw)$ of angle $2\theta$, with edges from $v$
given by $e^{i\alpha}$ and $e^{i(\alpha+2\theta)}$, a short calculation
yields
$$G(v,w)=-\frac{\theta}{\pi\tan\theta},$$
so that the part of 
$\partial{\cal E}/\partial\theta$ involving
the rhombus $R$ is $1/|\G_1|$ times
\begin{eqnarray*}
-2G(v,w)\frac{\partial(\Delta(v,w))}{\partial\theta}-
\frac{\partial}{\partial\theta}\left(\frac{2\theta}{\pi}\log\tan\theta\right)
&=&
\frac{2\theta}{\pi\tan\theta}\cdot\frac{\partial\tan\theta}{\partial\theta}-
\left(\frac{2\log\tan\theta}{\pi}+\frac{2\theta}{\pi\tan\theta\cos^2\theta}\right)\\
&=&-\frac{2\log\tan\theta}{\pi}.
\end{eqnarray*}

The integral from $0$ to $\phi$ is 
$$-\frac2{\pi}\int_0^{\phi}\log\tan\phi d\phi
=\frac2{\pi}(L(\phi)+L(\pi/2-\phi)),$$
where we used $L(\pi/2)=0$. 
Similarly the integral from $\pi/2$ down to $\phi$ is
$$\frac2{\pi}(L(\phi)+L(\pi/2-\phi)),$$
using $L(0)=L(\pi/2)=0$.
\hspace*{\fill} $\square$\medskip 

\section{Convexity}\label{cvxscn}

\begin{prop} When parametrized by the rhombus angles
the set of isoradial embeddings of a graph $\G$ is convex.
\end{prop}

\begin{proof}
Each rhombus has angle in $[0,\pi]$. The set of allowed
angles is the subset of $[0,\pi]^{|E|}$ defined by the linear
constraints that the sum of the angles around each vertex be $2\pi$.
It is therefore convex.
\end{proof}

\begin{thm}
A periodic graph which has an isoradial embedding
has a unique critical weight function $\nu$ maximizing $\det_1\bp,$
and a unique critical weight function $c$
maximizing $\det_1\Delta$.
\end{thm}
\begin{proof}
Given that the set of critical weight functions is convex,
and $Z:=\det_1\bp$ or $\det_1\Delta$ is continuous as a function of
the rhombus angles, we know that $Z$ attains its maximum.
To show unicity, 
it suffices to show that $\log Z$ is a strictly concave function
of the rhombus angles. Suppose there are $n$ rhombi in a fundamental
domain. Then $\log Z$ is the sum
over the rhombi in the fundamental domain of a function depending
only on the rhombus angle. Moreover, the functions
$$f_D(\theta)=\frac1{\pi}L(\theta)+\frac{\theta}{\pi}\log2\sin\theta$$
and
$$f_T(\theta)=\frac2{\pi}\left(L(\theta)+L(\frac{\pi}2-\theta)\right)+
\frac{2\theta}{\pi}\log\tan\theta$$
are strictly concave as a function of $\theta$. 
So we can view $\log Z$ as a function on $[0,\pi]^n$
(the set of rhombus angles),
subject to the linear restrictions that the sum of the angles
around each vertex
be $2\pi$. Thus $\log Z$ is the restriction to a linear subspace
of a strictly concave function on $[0,\pi]^n$. 
As such it is strictly concave and so has a unique maximum.
\end{proof}

What are the geometric intepretations of these maxima?

\section{Discretization}\label{discretscn}
Recall that a special discrete analytic function is a function
$f_b(z)$ satisfying (\ref{propagate1}) and (\ref{propagate2}).
Note that it is defined by its value at any vertex.
Other discret analytic functions can be built up from these
by convolutions:
\begin{thm}\label{specialdafs}
Let $f_b(z)$ be a special discrete analytic function and
$\mu$ be any measure in the plane. Then
the function $F(b) = \int f_b(z)d\mu(z)$
is a discrete analytic function.
\end{thm}

\begin{proof}
The proof was already given, in the proof of Theorem \ref{CF};
we repeat it here.
Let $w$ be a white vertex with neighbors $b_1,\dots,b_k$,
so that $R(wb_j)$ has edges $e^{i\theta_j}$ and $e^{i\phi_j}$.

We have 
\begin{eqnarray*}\sum_{b\in B}\bp(w,b)F(b)
&=& \sum_{j=1}^ki(-e^{i\theta_j}+e^{i\phi_j})F(b_j)\\
&=&\sum_{j=1}^ki(-e^{i\theta_j}+e^{i\phi_j})
\int f_{b_j}(z)d\mu(z)\\
&=&\int f_{w}(z) \sum_{j=1}^k
\frac{(-e^{i\theta_j}+e^{i\phi_j})}{(z-e^{i\theta_j})(z-e^{i\phi_j})}d\mu(z)\\
&=&\int f_{w}(z)\left(\sum_{j=1}^k
\frac1{(z-e^{i\phi_j})}-\frac1{(z-e^{i\theta_j})}\right)d\mu(z)\\
&=& \int 0 d\mu(z) = 0.
\end{eqnarray*}
\end{proof}

There is corresponding version for discrete harmonic functions,
which also appear in a different form in \cite{merc2}:
\begin{cor}
On $\G_T$, let $h_{v_0}(z)$ be given
and if $h_v(z)$ is defined and $v'$ is adjacent to $v$, define
$$h_{v'}(z)=h_v(z)\frac{(z+e^{i\theta})(z+e^{i\phi})}{
(z-e^{i\theta})(z-e^{i\phi})},$$ 
where $\theta$ and $\phi$ are the angles of the edges of rhombus
$R(vv')$.
This defines $h_v(z)$ for all $v\in \G_T$, and
$H(v)=\int h_v(z)d\mu(z)$ is discrete harmonic.
\end{cor}

\section{The dimer model}\label{dimerscn}
\subsection{Definitions}\label{dsdefs}
The dimer model on planar graphs was initiated by
Fisher, Kasteleyn and Temperley \cite{Kast1,FT}, who ``solved" the model
on $\Z^2$ by finding a closed-form expression for the partition function.

A dimer covering, or perfect matching, of a graph $\G$
is a set of edges $M$ of $\G$ which covers every vertex exactly once,
that is every vertex is an endpoint of exactly one element of $M$.
Let $\nu\colon E\to[0,\infty)$ be a weight function on the edges of $\G$.
To an edge $e$ we associate an energy $-\log\nu(e)$;
this defines a natural probability measure $\mu$, the Boltzman measure, on the
set of all dimer coverings $\cal M$, where the probability of a configuration
is proportional to the product of its dimer weights
(the exponent of minus the sum of the energies), 
$\mu(C)=\prod_{e\in C}\nu(e)$.
The {\bf dimer model} is the study of this measure.

The {\bf dimer partition function} is by definition
$${\cal Z}(\G,\nu) = \sum_{M\in\cal M} \nu(M).$$
The {\bf partition function per site} is by definition $Z
= {\cal Z}(\G,\nu)^{1/|\G|}$,
where $|\G|$ is the number of vertices of $\G$.

For infinite graphs one
can define the partition function per site by taking limits
of the partition functions per site on finite pieces:
if $\{\G_n\}$ is an exhaustion of $\G$ by finite subgraphs, one can define
$$Z=\lim_{n\to\infty} {\cal Z}(\G_n,\nu)^{1/|\G_n|},$$
when this limit exists. Usually this limit will depend on the sequence
$\{\G_n\}$: see e.g. \cite{CKP}. 

One can also define a probability measure 
$\mu$ on ${\cal M}(\G)$ which is a weak
limit $\mu=\lim_{n\to\infty} \mu_n$ where $\mu_n$ is the Boltzmann measure
on ${\cal M}(\G_n)$, assuming this limit exists. This limit is even more
dependent on the sequence than the partition function per site.

One general situation where these limits are known to exist
is when $\G_D$ is of the form $\G_T\cup\G_T^*$ as described in
section \ref{Lapscn}. For a periodic $\G_T$ the Green's functions
are known to converge to the corresponding Green's function on the plane
as defined in section \ref{greenscn}, see \cite{BP}.
From corollary \ref{K=G-G} we can compute the $\bp^{-1}$ operator
from the Green's function. In fact in this situation one can exhibit
a measure-preserving isomorphism from the dimer model on 
$\G_D$ to the corresponding
spanning tree model on $\G_T$ \cite{Temp,KPW}.
In particular the partitions functions are equal.

The utility of the $\bp$ operator in the dimer model is the
subject of the next section.
\subsection{Dimers on finite graphs}
A subgraph $\G_1$ of $\G$ is said to be 
{\bf simply connected} if it consists of
the set of edges and vertices contained on or inside a simple polygonal
path in $\G$.
\begin{thm}
Let $\G_1$ be a simply connected subgraph of a graph $\G$ (where $\G$
has a critical weight function), and $\bp_1$ the
submatrix of $\bp$ corresponding to $\G_1$.
The dimer partition function on $\G_1$ is $Z_1=\sqrt{|\det \bp_1|}$.
The probability of edges $\{w_1b_1,\dots,w_kb_k\}$ occurring
in a configuration chosen with respect to the Boltzmann measure $\mu$ is
$$\left(\prod_{i=1}^k \bp_1(w_i,b_i)\right) \det_{1\leq i,j\leq k}
((\bp_1)^{-1}(w_i,b_j)).$$
\end{thm}

The function $\bp_1^{-1}$ is called the {\bf coupling function} 
or {\bf inverse Kasteleyn matrix}.

\begin{proof}
For the first statement it suffices to show that $\bp$ is Kasteleyn-flat
\cite{Kuper}.
That is, we must show that for each face of $\G_1$ (except the outer face) 
with
vertices $u_1,v_1,\dots,u_m,v_m$ in cyclic order, we have
$$\arg(\bp(u_1,v_1)\dots \bp(u_m,v_m))=\arg(
(-1)^{m-1}\bp(v_1,u_2)\dots \bp(v_{m-1},u_m)\bp(v_m,u_1)).$$
(This identity implies that two dimer configurations which only differ
around a single face have the same argument in the expansion of
the determinant.
By \cite{Kuper}, any two configurations can be obtained from one another
by such displacements.)
To prove this identity, note that it is true if the points are
regularly spaced along a $2m$-gon, and note that it remains
true if you move one point at a time. 

For the second statement, see \cite{localstats}.
\end{proof}

\subsection{Volume and mean curvature of $P(\G)$}
\begin{prop}[Milnor \cite{Thu}] The volume of a hyperbolic simplex 
in the upper half space model with vertices at $\infty, (0,0,1),
(1,0,0)$ and $(\cos\theta,\sin\theta,0)$
is $L((\pi-\theta)/2)$, where $L$ is the Lobachevsky function (\ref{Lob}).
\end{prop}

Let $\G$ be a bipartite planar graph with an isoradial embedding,
and $P(\G^*)$ the hyperbolic polyhedron associated to its dual
(vertices of $\P$ are vertices of $\G^*$).
For each rhombus $R(wb)$ of $\G$ let $w,x,b,y$ be its four vertices;
we can decompose $P$ into simplices, so that the part of $P$ 
over $R$ consists of two simplices with vertices $w,x,y,\infty$
and $b,x,y,\infty$. Here vertices $x,y$ are ideal whereas
$w$ and $b$ are in the interior of hyperbolic three-space
(at Euclidean distance $1$ from the $xy$-plane).
Each of these simplices has volume 
$L(\theta)$, where $\theta$ is the half-angle of the rhombus.
Therefore in the expression for the determinant of $\bp$,
the sum of the terms involving $L$ is $\frac1{2\pi}$ times the volume of $P$.

The mean curvature of a compact convex polyhedron in $\H^3$
is by definition the sum over the edges of $(\pi-\theta_{\text{dih}})\ell$, 
where $\theta_{\text{dih}}$ is the dihedral angle and $\ell$ 
is the hyperbolic edge length.
For an ideal polyhedron, the edge lengths are infinite, so one first
assigns a small horosphere to each vertex and then computes the
edge lengths in the exterior of these horospheres. This definition
of course depends on the radii of the horospheres.
In the case of a $P$ arising from an isoradial embedding of $\G$
we can choose for a fixed small $\eps$ all the horospheres
to be Euclidean balls of diameter $\eps$;  in this case the
hyperbolic length of an edge between points $v_1$ and $v_2$
is exactly
$$2\log\frac{|v_2-v_1|}{\eps},$$
where $|v_2-v_1|$ is the {\it Euclidean} distance from $v_1$ to $v_2$.

For a rhombus of half-angle $\theta$, the dihedral angle of $P$
at the dual edge is $\pi-2\theta$.
Therefore the mean curvature per vertex of $P$ is 
\begin{equation}\label{mc}
\frac{4}{|\G_1|}\sum_{e\in\G_1}\theta\log\frac{2\sin\theta}{\eps}=
-4\pi\log\eps+\frac{4}{|\G_1|}\sum_{e\in\G_1}\theta\log2\sin\theta
\end{equation}
this last equality holds since at each vertex the sum of the 
half-angles is $\pi$. Finally it makes sense to
define the {\bf normalized mean curvature}
to be the above formula (\ref{mc}) without the $-4\pi\log\eps$ term.

\subsection{Geometry and Statistical mechanics}
If $\G_D$ arises as $\G_T\cup \G_T^*$ for a periodic isoradial
graph $\G_T$, we can identify
$\det_1\bp$ with the partition function of the associated
dimer model, and $\bp^{-1}$ as the coupling function, as
discussed at the end of section \ref{dsdefs}.

Moreover we have
\begin{thm} The volume per site of $P(\G_D^*)$ is $2\pi$ times
the entropy per site of the dimer
model. The normalized mean curvature per site of $P(\G^*_D)$ is $4\pi$ times
the mean energy per site of a dimer configuration.
\end{thm}

\begin{proof}
In the dimer model the probability of occurrence of an edge
$e=wb$ is $\nu(e)|\bp^{-1}(w,b)| = \frac{\theta}{\pi}$. 
Recall that the energy contribution for this edge is 
$\log2\sin\theta.$
The average energy per vertex $\bar E$ is therefore
$1/|\G_1|$ times the sum over the edges
in a fundamental domain of $\frac{\theta}\pi\log2\sin\theta$.
By the argument of Proposition 9.1 of \cite{CKP},
the fact that the edge correlations tend to zero implies that
the entropy per vertex of the dimer model is precisely 
$(\log \det_1\bp)- \bar E$, or

$$\frac1{|\G_1|}\sum_e \frac1{\pi}L(\theta),$$ 
that is, $1/2\pi$ times the volume per vertex of $P$. 
\end{proof}

\end{document}